\documentclass[aps,prd,
reprint]{revtex4-2}
\usepackage[LGR,T1]{fontenc}
\usepackage[latin9]{inputenc}
\usepackage{graphicx}
\usepackage{color}
\usepackage{textcomp}
\usepackage{dsfont}
\usepackage{amsmath}
\usepackage{esint}

\makeatletter

\usepackage{xfrac}

\date{\today}

\makeatother

\usepackage{babel}
\begin{document}
\title{A symmetry perspective of the Riemann zeros}
\author{Pushpa Kalauni$^{a}$ }
\email{pushpakalauni60@gmail.com (correspinding author)}

\author{Prasanta K. Panigrahi$^{b}$}
\email{pprasanta@iiserkol.ac.in }

\affiliation{$^{a}$Department of Physics, Indian Institute of Space Science and
Technology, Thiruvananthapuram, Kerala, 695547, India~\\
 $\ensuremath{^{b}}$Department of Physical Sciences, Indian Institute
of Science Education and Research Kolkata, Mohanpur 741246, West Bengal,
India}
\begin{abstract}
We study the relationship between the zeros of the Riemann zeta function
and physical systems exhibiting supersymmetry, $PT$ symmetry and
$SU(2)$ group symmetry. Our findings demonstrate that unbroken supersymmetry
is associated with the presence of non-trivial zeros of the zeta function. However, in other cases, supersymmetry is spontaneously broken and
the ground state energy of the system is not zero. Moreover, we have
established the manifestation of PT symmetry invariance within our
supersymmetric system. In addition, our findings
provide insights into a $SU(2)$ symmetry that arises within these
systems, with the Hilbert space having a two-level structure.
\end{abstract}
\maketitle

\section{Introduction}

Supersymmetric quantum mechanics (SUSY QM) \cite{witten,cooper,das1}
is a theoretical framework that combines ideas from quantum mechanics
and supersymmetry, a symmetry that relates particles with different
spin values. One of the interesting features of SUSY QM is that its
ground state energy always vanishes, a consequence of the supersymmetry
of the theory.

Recently, there has been a fascinating development in the study of
SUSY QM, which connects it to the Riemann Hypothesis \cite{Riemann},
a famous problem in mathematics that deals with the distribution of
prime numbers \cite{kalauni}. Specifically, it has been shown that
the non-trivial zeros of the Riemann zeta function can be described
as the ground state energy of certain SUSY QM models. 

In some recent studies, the relationship between the Riemann zeta
function and various aspects of Physics has been explored \cite{schumayer,wolf,bender,savvidy,mcguigan,murugesh}.
In another study, the Riemann zeta function relates to scattering
amplitudes in quantum field theory \cite{remmen}. Additionally, the
Riemann zeta function and its relevance to second-order supersymmetric
quantum mechanics have been discussed \cite{garcia}. \\
The Riemann zeta function \cite{edwards} is defined for complex values
of $s$ in the following form as
\begin{align}
\zeta(s)= & \frac{1}{(1-2^{1-s})}\sum_{n=1}^{\infty}(-1)^{n+1}n^{-s},\,\,\,\Re(s)>0.\label{eq:1}
\end{align}
The Riemann zeta function can also be defined in the complex plane
by the contour integral \cite{edwards}
\begin{align}
\zeta(s) & =\frac{\Gamma(1-s)}{2\pi i}\int_{C}\frac{t^{s-1}}{e^{-t}-1}dt,\label{eq:2}
\end{align}
where the contour of integration encloses the negative $t$-axis,
looping from $t=-\infty-i0$ to $t=-\infty+i0$ enclosing the point
$t=0$. 

The Riemann zeta function possesses both trivial and non-trivial zeros.
Trivial zeros of the zeta function occur when $s$ is negative even
integer. These zeros can be obtained from the functional relation
of the Riemann zeta function, which is given by
\begin{align}
\zeta(s) & =2^{s}\pi^{s-1}\sin(\frac{\pi s}{2})\Gamma(1-s)\zeta(1-s),\,s<1.\label{eq:3}
\end{align}
Non-trivial zeros are associated with the Riemann hypothesis, which
asserts that all non-trivial zeros of the Riemann zeta function $\zeta(s)$
lie on the critical line, the line in the complex plane where the
real part of $s$ is equal to $1/2$. To incorporate trivial zeros
into the supersymmetric model, a modified inner product was introduced,
and the Hilbert space of the supersymmetric system was defined by
implementing appropriate boundary conditions. This was discussed in
more detail in a study by Kalauni et al. \cite{milton}.

In \cite{milton}, we define a supersymmetric model in which the eigenvalue
of the system arises in the form of $\zeta(s)\zeta(1-s)$ (where $s=\sigma+i\omega$,
$\sigma$ and $\omega$ are real), which is the product of the Riemann
zeta function evaluated at complex arguments $(\sigma+i\omega)$ and
$(1-\sigma-i\omega)$. It was demonstrated that the states of the
supersymmetric system under consideration form an orthonormal set
of functions, when the Hilbert space is defined for a finite interval
of real line, specifically in $L^{2}(1,a)$ with parameters $\sigma=1/2$
and $a=e^{2\pi/\omega}$. As a result, it was shown that the ground
state energy vanishes only when $\sigma=1/2$. However, it is natural
to ask whether there are any additional indications which suggest
that the ground state energy vanishes only when $\sigma=1/2$. The
aim of this paper is to gain a more comprehensive understanding of
the supersymmetric system under consideration by exploring different
symmetries within the system. To achieve this, we employ various methods,
including analyzing the Witten index, examining $PT$ symmetry, and
investigating  $SU(2)$ symmetry present in the system.

In the next section, we provide a review of the SUSY QM model, which
yields the Riemann zeta function as an eigenvalue of the system. In
the third section, the Witten index is used to examine the unbroken
and broken supersymmetry of the system. It is shown that for $\sigma=1/2$,
it gives real eigenvalues of the system while for other values of
$\sigma$, the energy of the system becomes complex, indicating a
possible connection with $PT$ symmetry \cite{bender1,bender2}.

$PT$ symmetry is a concept in quantum mechanics which refers to the
behavior of physical systems that are invariant under the combined
operations of parity $(P)$ and time reversal $(T)$. One of the interesting
aspects of $PT$ symmetry is that it can lead to the existence of
complex eigenvalues and non-Hermitian operators in quantum mechanics.
In the fourth section, we investigate the $PT$ symmetry in our supersymmetric
system and establish that it remains unbroken only if a specific condition
holds. However, for other cases, PT symmetry is broken. In the fifth
section, we show that this Hilbert space of the system forms a $SU(2)$
symmetry (function-dependent) which is present in the system because
of the presence of the trivial and the non-trivial zeros of the zeta
function. 

These results provide new insights into the underlying algebraic structures
and symmetries of the system. This gives the motivation for comprehending
the interrelations between the Riemann zeros and diverse symmetries,
namely supersymmetry, $PT$ symmetry, and $SU(2)$ symmetry.

\section{Supersymmetric System and Riemann zeta function}

To begin, we review our supersymmetric model \cite{milton}, which
yields the Riemann zeta function as an eigenvalue. This supersymmetric
model is defined by introducing lowering and raising operators (denoted
by $A$ and $A^{\dagger}$, respectively), which is given as
\begin{align}
A & =x^{-i\omega}\Omega,\nonumber \\
A^{\dagger} & =\Omega^{\dagger}x^{i\omega},\label{eq:4}
\end{align}
where 
\begin{align}
\Omega= & \frac{\Gamma(x\frac{d}{dx}+1)}{2\pi i}\intop_{C}\frac{t^{-x\frac{d}{dx}-1}}{e^{-t}-1}dt,\label{eq:5}
\end{align}
and $\omega$ is real and contour $C$ is as given in Eq.~(\ref{eq:2}).
When the operator $\Omega$ (defined in Eq.~(\ref{eq:5})) acts on
the monomial $x^{-s}$, it produces the eigenvalues in terms of the
Riemann zeta function as \cite{brody}
\begin{align}
\Omega x^{-s} & =\left[\frac{\Gamma(x\frac{d}{dx}+1)}{2\pi i}\intop_{C}\frac{t^{-x\frac{d}{dx}-1}}{e^{-t}-1}dt\right]x^{-s},\nonumber \\
 & =\left[\frac{\Gamma(1-s)}{2\pi i}\intop_{C}\frac{t^{s-1}}{e^{-t}-1}dt\right]x^{-s},\nonumber \\
 & =\zeta(s)\,x^{-s}.\label{eq:6}
\end{align}
Using these operators $\Omega$ and $\Omega^{\dagger}$, it is possible
to construct a set of eigenstates for the Hamiltonian of the system,
which correspond to different energy levels of the system. These eigenstates
can be used to compute the spectrum of the system and determine its
ground state energy.

The Hamiltonian of the supersymmetric system can be written in terms
of a $2\times2$ matrix as
\begin{align}
H= & \left[\begin{array}{cc}
H_{-} & 0\\
0 & H_{+}
\end{array}\right],\label{eq:7}
\end{align}
where 
\begin{align}
H_{-} & =A^{\dagger}A=\Omega^{\dagger}\Omega,\nonumber \\
H_{+} & =AA^{\dagger}=x^{-i\omega}\Omega\Omega^{\dagger}x^{i\omega},\label{eq:8}
\end{align}
are supersymmetric partner Hamiltonians.

We define the wavefunction as $x^{-s}$ with $s=\sigma+i\omega$ (where
$\sigma$ and $\omega$ are real). It gives eigenvalues of the system
in terms of the Riemann zeta function as
\begin{align}
H_{-}x^{-\sigma-i\omega} & =\zeta(\sigma+i\omega)\zeta(1-\sigma-i\omega)x^{-\sigma-i\omega}.\label{eq:9}
\end{align}
Eq.~(\ref{eq:9}) shows that $H_{-}$ has real eigenvalues only if
$\sigma=1/2$. This means that when $\sigma\neq1/2$, then the eigenvalues
of $H_{-}$ will be complex numbers.

In our specific model \cite{kalauni,milton}, the ground state energy
of the system is described in terms of the Riemann zeta function.
Our analysis indicates that the unbroken supersymmetry present in
the system results as a vanishing ground state energy of the system.
We elaborate on the concept of unbroken supersymmetry and its relation
to the ground state energy in the subsequent section by utilizing
the Witten index.

\section{Unbroken and Broken Supersymmetry}

The Witten index is a powerful tool for studying supersymmetric theories.
One of the most important applications of the Witten index is in determining
whether or not supersymmetry is broken in a given theory. 

To calculate the Witten index for a given system, we count the number
of bosonic and fermionic states with zero energy, denoted as $n_{B}^{E=0}$
and $n_{F}^{E=0}$, respectively. These counts are done under three
distinct conditions, which Witten identified in his paper \cite{witten}:
\begin{enumerate}
\item If $n_{B}^{E=0}-n_{F}^{E=0}\neq0$, then supersymmetry is unbroken.
\item If $n_{B}^{E=0}=n_{F}^{E=0}=0$, supersymmetry is spontaneously broken.
\item If $n_{B}^{E=0}$ and $n_{F}^{E=0}$ are equal but non-zero, supersymmetry
is unbroken.
\end{enumerate}
We start by defining the wavefunction as $x^{-s}$ with $s=\sigma+i\omega$
(where $\sigma$ and $\omega$ are real) and for applying Witten index
condition, we check that 
\begin{align}
H_{-}x^{-s} & =\zeta(s)\zeta(1-s)x^{-s};\nonumber \\
H_{+}x^{s-1-i\omega} & =\zeta(s)\zeta(1-s)x^{s-1-i\omega},\label{eq:10}
\end{align}
which shows that when the parameter $s$ is fixed, there exists a
bosonic state denoted as $x^{-s}$ and a corresponding fermionic state
denoted as $x^{s-1-i\omega}$, where both states contribute the same
amount of energy. However, it is important to note that the bosonic
and fermionic states are normalized and form a Hilbert space if $\text{Re}(s)=1/2$.
As we see that if $\text{Re}(s)=1/2$, $\text{Im}(s)=\omega$, and
$\ensuremath{\zeta(1/2+i\omega)=0}$, we find a bosonic state $x^{-1/2+i\omega}$
and a corresponding fermionic state $x^{-1/2-i\omega}$, both of which
contribute zero energy to the ground state. This fulfills the third
condition of the Witten index, where both $n_{B}^{E=0}$ and $n_{F}^{E=0}$
are equal to $1$, resulting in zero difference and indicating unbroken
symmetry. However, if $\text{Re}(s)=1/2$ and $\omega$ does not correspond
to the imaginary part of the non-trivial zeros of the zeta function,
then supersymmetry remains broken due to Witten's second condition.

For values of $\text{Re(s)}$ other than $1/2$, no solutions exist
within the Hilbert space, resulting in both $n_{B}^{E=0}$ and $n_{F}^{E=0}$
being zero, which indicates symmetry breaking. This implies that when
$\text{Re(s)\ensuremath{\neq}1/2}$, then the eigenvalues of $H_{-}$
are complex numbers. 

This peculiarity of complex eigenvalues can be linked to the $PT$
symmetry of the system. In the following section, we elaborate on
the $PT$ symmetry of the system and its relation to the eigenvalues
of the Hamiltonian. 

\section{$PT$ Unbroken and Broken Symmetry}

In \cite{milton}, it has been demonstrated that the $n^{th}$ states
of the system are normalized and constitute a complete set of functions
in the interval $[1,a]$ of square-integrable functions $L^{2}$.
The value of $a$ is related to non-trivial zeros of the zeta function,
specifically, if $\zeta(\frac{1}{2}+i\omega)=0$ where $\omega=2\pi/\log a$.
Additionally, these functions are also orthonormal in the interval
$[-a,-1]$ of square-integrable functions $L^{2}$.

In order to define $PT$ symmetry, it is necessary to have a Hilbert
space where the operations $x\rightarrow-x$ and $p\rightarrow p$
can be utilized. It indicates that we can use $PT$
symmetry for this particular type of function, which may have important
implications for the properties and behavior of the system.

To incorporate both positive and negative values of $x$, we take
the direct sum of these two Hilbert spaces denoted as $L^{2}[-a,-1]\oplus L^{2}[1,a]$.
Therefore, we define wave functions in the following form
\begin{align}
\psi_{1}(x) & =|x|^{-\sigma-i\omega}\,\,\,\,\,\,\,\,\,\,\,\,\,\psi_{2}(x)=\text{sgn}(x)|x|^{-\sigma-i\omega}.\label{eq:11}
\end{align}
If the Hamiltonian $H_{-}$ is invarinat under $PT$ symmetry, it
satisfies,
\begin{align}
[PT,H_{-}]\psi_{1}(x) & =0,\label{eq:12}
\end{align}
It follows that 
\begin{align}
\zeta(\sigma+i\omega)\zeta(1-\sigma-i\omega) & -\zeta(\sigma-i\omega)\zeta(1-\sigma+i\omega)=0.\label{eq:13}
\end{align}
Similarly, we can check $PT$ symmetry for $\psi_{2}(x)$ as
\begin{align}
[PT,H_{-}]\psi_{2}(x) & =0,\label{eq:14}
\end{align}
which is true if Eq.~(\ref{eq:13}) is satisfied.

An important insight derived from this connection is that $PT$ symmetry
in the system remains invariant in two cases: i) $\text{Re}(s)=\sfrac{1}{2}$
and ii) $\text{Im}(s)=0$ (where $s=\sigma+i\omega)$.

This finding implies that $PT$ symmetry, similar to supersymmetry,
remains unchanged when the $\text{Re}(s)=\sfrac{1}{2}$. However,
for values of $s$ other than those specified, the $PT$ symmetry
of the system is broken.\\

This system also possesses an underlying $SU(2)$
algebraic structure like Quasi-exactly solvable system \cite{ushveridze}.
In the following section, we demonstrate this wavefunction-dependent
$SU(2)$ symmetry, which arises subject to the same constraint outlined
in Eq.~(\ref{eq:13}). 

\section{$SU(2)$ symmetry}

$SU(2)$ is a Lie group that describes the group of special unitary
transformations. The generators of $SU(2)$ are the angular momentum
operators $J_{-}$, $J_{0}$, and $J_{+}$, which satisfy the following
commutation relations
\begin{align}
[J_{+},J_{-}] & =2J_{0},\nonumber \\{}
[J_{0},J_{\pm}] & =\pm J_{\pm}.\label{eq:15}
\end{align}
To establish the link between the Riemann zeros and $SU(2)$ symmetry,
we begin by defining $J_{-}$ and $J_{+}$ using the lowering and
raising operators of our supersymmetric system, with a normalizing
factor of $1/\sqrt{\zeta(k)\zeta(1-k)}$:
\begin{align}
J_{-}=\frac{A}{\sqrt{\zeta(k)\zeta(1-k)}}= & \frac{1}{\sqrt{\zeta(k)\zeta(1-k)}}x^{-i\omega}\Omega,\nonumber \\
J_{+}=\frac{A^{\dagger}}{\sqrt{\zeta(k)\zeta(1-k)}}= & \frac{1}{\sqrt{\zeta(k)\zeta(1-k)}}\Omega^{\dagger}x^{i\omega},\label{eq:16}
\end{align}
 and we chose values of $k=1/2$ for non-trivial zeros and $k=-2N-i\omega$
for the case of trivial zeros. \\
The commutation relation between $J_{+}$ and $J_{-}$ leads to,
\begin{align}
[J_{+},J_{-}]=\frac{(\Omega^{\dagger}\Omega-x^{-i\omega}\Omega\Omega^{\dagger}x^{i\omega})}{\zeta(k)\zeta(1-k)} & =2J_{0},\label{eq:17}
\end{align}
where
\begin{align}
J_{0} & =\frac{(\Omega^{\dagger}\Omega-x^{-i\omega}\Omega^{\dagger}\Omega x^{i\omega})}{2\zeta(k)\zeta(1-k)}.\label{eq:18}
\end{align}
We define $J_{x}$ and $J_{y}$ in the following form,
\begin{align}
J_{x} & =\frac{\Omega^{\dagger}x^{i\omega}+x^{-i\omega}\Omega}{2\sqrt{\zeta(k)\zeta(1-k)}},\nonumber \\
J_{y} & =\frac{\Omega^{\dagger}x^{i\omega}-x^{-i\omega}\Omega}{2i\sqrt{\zeta(k)\zeta(1-k)}}.\label{eq:19}
\end{align}
The action of the generators $J_{x}$, $J_{y}$, and $J_{0}$ on a
function $x^{-s}$ yields,
\begin{align}
J_{x}x^{-s} & =\frac{\zeta(1-s+i\omega)x^{-s+i\omega}+\zeta(s)x^{-s-i\omega}}{2\sqrt{\zeta(k)\zeta(1-k)}},\nonumber \\
J_{y}x^{-s} & =\frac{\zeta(1-s+i\omega)x^{-s+i\omega}-\zeta(s)x^{-s-i\omega}}{2i\sqrt{\zeta(k)\zeta(1-k)}},\nonumber \\
J_{0}x^{-s} & =\left[\frac{\zeta(s)\zeta(1-s)-\zeta(s-i\omega)\zeta(1-s+i\omega)}{2\zeta(k)\zeta(1-k)}\right]x^{-s}.\label{eq:20}
\end{align}
For $SU(2)$, the Casimir operator is given by,
\begin{align}
J^{2} & =J_{x}^{2}+J_{y}^{2}+J_{0}^{2},\label{eq:21}
\end{align}
that commutes with each of these generators $J_{x}$, $J_{y}$ and
$J_{0}$. \\
The Casimir operator acts on the function $x^{-s}$ as
\begin{align}
J^{2}x^{-s}= & \left[\frac{\zeta(s-i\omega)\zeta(1-s+i\omega)+\zeta(s)\zeta(1-s)}{2\zeta(k)\zeta(1-k)}\right]x^{-s}\nonumber \\
 & +\left[\frac{\zeta(s-i\omega)\zeta(1-s+i\omega)-\zeta(s)\zeta(1-s)}{2\zeta(k)\zeta(1-k)}\right]^{2}x^{-s},\label{eq:22}
\end{align}
\textcolor{black}{where $s=\sigma+i\omega$. The structures of the
commutators and the Casimir operator on this Hilbert space are indicative
of a Hilbert-space-dependent structure \cite{schiff}. }

We will now demonstrate how the $SU(2)$ algebra is related to the
existence of both non-trivial and trivial zeros in zeta functions,
possessing a Hilbert-space-dependent spin $1/2$ structure. 

\subsection{Non-trivial Riemann zeros}

We define a $SU(2)$\textbf{\textcolor{red}{{} }}\textcolor{black}{eigenstate}
as
\begin{align}
|\sigma,m\rangle & =\frac{1}{\sqrt{\log a}}x^{-\sigma+i(m-\frac{1}{2})\omega},\label{eq:23}
\end{align}
where $a=e^{2\pi/\omega}.$ The spin of a particle with quantum number
$s$ can take $2s+1$ values for the unitary irreducible representation.
\textcolor{black}{In the present case, the Hilbert space has a two-level
$\sigma=1/2$ representation}\textbf{\textcolor{red}{.}}\textbf{ }These
two states can be represented as $|\sigma,m\rangle=$$|1/2,1/2\rangle$
and $|1/2,-1/2\rangle$. When $\sigma=1/2$, these states form a discrete,
orthonormal, and complete basis in a finite interval of real line
\cite{milton} and the orthonormality condition for this basis is
given by
\begin{align}
\langle1/2,m|1/2,m'\rangle & =\delta_{mm'}.\label{eq:24}
\end{align}
We assume that $\omega$ is the imaginary part of non-trivial zeros
of the zeta function and define
\begin{align}
\zeta(\frac{1}{2}+i\omega)=\zeta(\frac{1}{2}-i\omega) & =0.\label{eq:25}
\end{align}
Using $k=1/2$ and Eq.~(\ref{eq:25}), we can write Eq.~(\ref{eq:20})
as 
\begin{align}
J_{x}|1/2,1/2\rangle & =\frac{1}{2}|1/2,-1/2\rangle,\nonumber \\
J_{y}|1/2,1/2\rangle & =-\frac{1}{2i}|1/2,-1/2\rangle,\nonumber \\
J_{0}|1/2,1/2\rangle & =\frac{1}{2}|1/2,1/2\rangle,\label{eq:26}
\end{align}
which provides the eigenvalues of $J{{}^2}$ and $J_{0}$ as
\begin{align}
J^{2}|1/2,\pm1/2\rangle & =\frac{3}{4}|1/2,\pm1/2\rangle,\nonumber \\
J_{0}|1/2,\pm1/2\rangle & =\pm\frac{1}{2}|1/2,\pm1/2\rangle.\label{eq:27}
\end{align}
From Eqs.~(\ref{eq:16}-\ref{eq:27}), we can see that the operators
$J_{+}$, $J_{-}$, and $J_{0}$ satisfy the well-known $SU(2)$ algebraic
commutation relations 
\begin{align}
[J_{+},J_{-}]|1/2,1/2\rangle & =(2J_{0})|1/2,1/2\rangle,\nonumber \\{}
[J_{0},J_{-}]|1/2,1/2\rangle & =-J_{-}|1/2,1/2\rangle,\nonumber \\{}
[J_{0},J_{+}]|1/2,-1/2\rangle & =J_{+}|1/2,-1/2\rangle.\label{eq:28}
\end{align}
It is noteworthy that $SU(2)$ algebra in this system is dependent
on the zeta function, and the zeta function must satisfy a condition
given in Eq. (\ref{eq:25}) in order to satisfy this algebra. This
constraint implies that $\sigma$ must be $1/2$, indicating that
it applies to the non-trivial zeros of the zeta function. We explain
the significance of the non-trivial zeros of the Riemann zeta function
for demonstrating the $SU(2)$ algebra. Now, we examine how the trivial
zeros help us understand the $SU(2)$ algebra in our system.

\subsection{Trivial Riemann zeros:}

It is well known that the Riemann zeta function $\zeta(\sigma+i\omega)$
possesses trivial zeros when $\sigma=-2N$ and $\omega=0$. This can
be understood through the functional relation of the zeta function,
which is given in Eq.~(3).

In this case, we define the state
\begin{align}
\psi_{\sigma,m} & =\frac{1}{\sqrt{\log a}}x^{-\sigma+i(\frac{1}{2}-m)\omega}.\label{eq:29}
\end{align}
These functions do not form the orthonormal set under Dirac inner
product unless $\sigma=1/2$, which is the case for non-trivial zeros.
For other values of $\sigma$, we have introduced the modified inner
product \cite{das2,milton} then it forms an orthonormal set of functions.

Using $\sigma=-2N$, $k=-2N-i\omega$, $m=1/2$ and using Eq.~(\ref{eq:29}),
one can write Eq.~(\ref{eq:20}) as,
\begin{align}
J_{x}\psi_{-2N,1/2} & =\frac{\zeta(1+2N+i\omega)}{2\sqrt{\zeta(-2N-i\omega)\zeta(1+2N+i\omega)}}\psi_{-2N,-1/2},\nonumber \\
J_{y}\psi_{-2N,1/2} & =\frac{\zeta(1+2N+i\omega)}{2i\sqrt{\zeta(-2N-i\omega)\zeta(1+2N+i\omega)}}\psi_{-2N,-1/2},\nonumber \\
J_{0}\psi_{-2N,1/2} & =-\frac{1}{2}\psi_{-2N,1/2}.\label{eq:30}
\end{align}
It leads to
\begin{align}
J^{2}=(J_{x}^{2}+J_{y}^{2}+J_{0}^{2}) & =\frac{3}{4},\label{eq:31}
\end{align}
 which also implies that $j=\frac{1}{2}$ and $m$ varies from $-1/2$
to $1/2$.

We see that to establish the functional-dependent $SU(2)$ algebra,
it is essential to have either non-trivial or trivial zeros of the
zeta function. Without the presence of either of these types of Riemann
zeros, our model would not exhibit the algebraic structure of $SU(2)$.
This indicates that the existence of either type of zero plays a critical
role in the formation of the $SU(2)$ algebra here, which is an appealing
outcome.

\section{Conclusion}

We have demonstrated that various symmetries are linked to both the
Riemann zeros and the supersymmetric system underlying it. Supersymmetry
naturally leads to the utilization of the Witten index for the characterization
of ground state energy. Interestingly, it is shown that SUSY is unbroken
for $\text{Re}(s)=\sigma=1/2$. The appearance of PT symmetry and
the function-dependent $SU(2)$ symmetries are deeply connected to
the properties of the Hilbert space, which can be characterized by
the Riemann zeros. Our results show that SUSY remains unbroken only
when $\sigma=1/2$, which corresponds to the non-trivial zeros of
the zeta function. We have also demonstrated that the $PT$ symmetry
remains unbroken in two cases: i) $\text{Re}(s)=\sfrac{1}{2}$ and
ii) $\text{Im}(s)=0$ (where $s=\sigma+i\omega)$ while in other cases,
the $PT$ symmetry is broken. We find an underlying
$SU(2)$ algebraic structure, where the algebra is realized in the
Hilbert space with suitable constraints. Intriguingly, only the spin-half
representation is realized. The Hilbert space-dependent
$SU(2)$ symmetry and the presence of only spin-half unitary irreducible
representation are indicative of a deeper structure, needing further
study.

Our findings provide novel perspectives on the Riemann zeros and their
association with various symmetries. These symmetries shed light on
the fundamental connections between the Riemann hypothesis and physics,
and highlight the potential for new insights and discoveries at the
intersection of these two fields.
\begin{acknowledgments}
We acknowledge Prof. Kimball A. Milton for the useful discussions.
This work is financially supported by the DST, Govt. of India under
the Women Scientist A, Ref. No. DST/WOS-A/PM-64/2019. We are thankful
to the referees for providing their valuable feedback.\\
\end{acknowledgments}

\appendix

\section{Derivation Details of Equations in Section ``SU(2) Symmetery''}

Using Eq.~(\ref{eq:16}) and Eq.~(\ref{eq:18}), we get
\begin{align}
[J_{+},J_{-}] & =\frac{1}{b(k)}(\Omega^{\dagger}\Omega-x^{-i\omega}\Omega\Omega^{\dagger}x^{i\omega}),\nonumber \\{}
[J_{0},J_{-}] & =\frac{1}{2(b(k))^{3/2}}[\Omega^{\dagger}\Omega x^{-i\omega}\Omega-x^{-i\omega}\Omega\Omega^{\dagger}\Omega\nonumber \\
 & -x^{-i\omega}\Omega\Omega^{\dagger}\Omega+x^{-i\omega}\Omega x^{-i\omega}\Omega\Omega^{\dagger}x^{i\omega}],\nonumber \\{}
[J_{0},J_{+}] & =\frac{1}{2(b(k))^{3/2}}[\Omega^{\dagger}\Omega\Omega^{\dagger}x^{i\omega}-x^{-i\omega}\Omega\Omega^{\dagger}x^{i\omega}\Omega^{\dagger}x^{i\omega}\nonumber \\
 & -\Omega^{\dagger}x^{i\omega}\Omega^{\dagger}\Omega+\Omega^{\dagger}\Omega\Omega^{\dagger}x^{i\omega}],\label{eq:32}
\end{align}
where 
\begin{align}
b(k) & =\zeta(k)\zeta(1-k).\label{eq:33}
\end{align}
The operator $\Omega$ and its adjoint $\Omega^{\dagger}$, operate
on monomials $x^{-s}$ and provide
\begin{align}
\Omega x^{-s} & =\zeta(s)x^{-s},\nonumber \\
\Omega^{\dagger}x^{-s} & =\zeta(1-s)x^{-s}.\label{eq:34}
\end{align}
We use Eq.~(\ref{eq:34}) and calculate the commutation relations
as
\begin{align}
[J_{+},J_{-}]x^{-s} & =\frac{[b(s)-b(s-i\omega)]}{b(k)}x^{-s},\nonumber \\{}
[J_{0},J_{-}]x^{-s} & =[b(s+i\omega)+b(s-i\omega)-2b(s)]\times\nonumber \\
 & \frac{\zeta(s)}{2(b(k))^{3/2}}x^{-s-i\omega},\nonumber \\{}
[J_{0},J_{+}]x^{-s-i\omega} & =[-b(s-i\omega)-b(s+i\omega)+2b(s)]\times\nonumber \\
 & \frac{\zeta(1-s)}{2(b(k))^{3/2}}x^{-s}.\label{eq:35}
\end{align}
$J_{x}^{2}$ and $J_{y}^{2}$ act on the monomial $x^{-s}$ as

\begin{align}
J_{x}^{2}x^{-s}= & \frac{\zeta(1-s+i\omega)}{4}\nonumber \\
\times & \left[\frac{\zeta(1-s+2i\omega)x^{-s+2i\omega}+\zeta(s-i\omega)x^{-s}}{b(k)}\right]\nonumber \\
+ & \frac{\zeta(s)}{4}\left[\frac{\zeta(1-s)x^{-s}+\zeta(s+i\omega)x^{-s-2i\omega}}{b(k)}\right],\nonumber \\
J_{y}^{2}x^{-s}= & \frac{-\zeta(1-s+i\omega)}{4}\nonumber \\
\times & \left[\frac{\zeta(1-s+2i\omega)x^{-s+2i\omega}-\zeta(s-i\omega)x^{-s}}{b(k)}\right]\nonumber \\
+ & \frac{\zeta(s)}{4}\left[\frac{\zeta(1-s)x^{-s}-\zeta(s+i\omega)x^{-s-2i\omega})}{b(k)}\right],\label{eq:36}
\end{align}
where $s=\sigma+i\omega$. 

\begin{thebibliography}{10}
\bibitem{witten} E. Witten, Constraints on Supersymmetry breaking,
Nucl. Phys. B 202 (1982) 253.

\bibitem{cooper} F. Cooper, A. Khare and U. Sukhatme, Supersymmetry
in Quantum Mechanics, World Scientific, Singapore, (2001)

\bibitem{das1} A. Das, Field theory: a path integral approach. Vol.
83. World Scientific, (2019).

\bibitem{Riemann} B. Riemann, Ueber die Anzahl der Primzahlen unter
einer gegebenen Grosse, Ges. Math. Werke und Wissenschaftlicher Nachla\ss,
2 (1859) 145.

\bibitem{kalauni} A. Das and P. Kalauni, Supersymmetry and the Riemann
zeros on the critical line, Phys. Lett. B 791 (2019) 265. 

\bibitem{schumayer} D. Schumayer and D. A. W Hutchinson, Physics
of the Riemann hypothesis, Rev. Mod. Phys. 83 (2011) 307.

\bibitem{wolf} M. Wolf, Will a physicist prove the Riemann hypothesis?
Rep. Prog. Phys. 83 (2020) 036001. 

\bibitem{bender} C. M. Bender, D. C. Brody and M. P. Muller, Hamiltonian
for the zeros of the Riemann zeta function, Phys. Rev. Letts. 118
(2017) 130201.

\bibitem{savvidy} G. Savvidy and K. Savvidy, Quantum-mechanical interpretation
of Riemann zeta function zeros, arXiv:1809.09491 (2018).

\bibitem{mcguigan} M. Mcguigan, Riemann hypothesis, modified Morse
potential and supersymmetric quantum mechanics, arXiv:2002.12825 (2020).

\bibitem{murugesh} P. Kalauni, A. Kumar, and S. Murugesh, Supersymmetric
quantum mechanical system for locating the Riemann zeros, Eur. Phys.
J. Plus, 138 (2023) 487.

\bibitem{remmen} G. N. Remmen, Amplitudes and the Riemann zeta function,
Phys, Rev. Letts. 127 (2021) 241602.

\bibitem{garcia} D. Garcia-Mu\~{n}oz Juan, A. Raya, and Y. Concha Sanchez,
Second-Order SUSY-QM and zeroes of the Riemann zeta function,  arXiv:2301.05360
(2023).

\bibitem{edwards}  H. M. Edwards, Riemann's zeta function, Academic
Press, New York, (1974).

\bibitem{milton} P. Kalauni, and K. A. Milton, Supersymmetric quantum
mechanics and the Riemann hypothesis, Int. J. Mod. Phys. A, (2023),
doi:10.1142/S0217751X23501105.

\bibitem{bender1} C. M. Bender and S. Boettcher, Real spectra in
non-Hermitian Hamiltonians having $PT$ symmetry., Phys. Rev. Lett.
80 (1998) 5243.

\bibitem{bender2} C. M. Bender, S. Boettcher, P. N. Meisinger, PT-symmetric
quantum mechanics, J. Math. Phys. 40 (1999) 2201. 

\bibitem{brody} C. M. Bender and D. C. Brody, Operator-valued zeta
functions and Fourier analysis, J. Phys. A: Math. Theor. 52 (2019)
345201.

\bibitem{ushveridze} A. G. Ushveridze, Quasi-exactly solvable models
in quantum mechanics. CRC Press; (1994). 

\bibitem{schiff} L. I. Schiff, Quantum Mechanics 3rd, M cGraw-Hill,
New York (1968).

\bibitem{das2} A. Das and P. Kalauni, Operator description for thermal
quantum field theories on an arbitrary path in the real time formalism,
Phys. Rev. D 93 (2016) 125029.
\end{thebibliography}
\end{document}